\documentclass[prl,twocolumn,amsmath,amssymb,superscriptaddress]{revtex4-2}
\usepackage{epsfig, graphicx,graphics,amsmath,amssymb,float}
\usepackage[T1]{fontenc}
\usepackage[latin9]{inputenc}
\usepackage{amsmath}
\usepackage{amssymb}
\usepackage{appendix}
\usepackage{amscd}
\usepackage{bm}
\usepackage{psfrag}
\usepackage{bbm} 
\usepackage{babel}
\usepackage{wasysym }
\usepackage{mathrsfs}
\usepackage{color}
\usepackage[normalem]{ulem}
\usepackage[final]{hyperref} 
\hypersetup{
	colorlinks=true,       
	linkcolor=blue,        
	citecolor=blue,        
	filecolor=magenta,     
	urlcolor=blue         
}
\usepackage{tikz}
\usetikzlibrary{automata, positioning, arrows}
\usepackage{qcircuit}
\usepackage{tikz}

\definecolor{darkblue}{HTML}{004D6B}
\definecolor{darkred}{HTML}{8c1515}
\definecolor{darkgreen}{HTML}{006400}

\newcommand{\ba}{\begin{array}}
\newcommand{\ea}{\end{array}}
\newcommand{\be}{\begin{equation}}
\newcommand{\ee}{\end{equation}}
\newcommand{\bea}{\begin{eqnarray}}
\newcommand{\eea}{\end{eqnarray}}

\usepackage[american,siunitx]{circuitikz}
\usepackage{amsmath}
\usetikzlibrary{shapes,decorations.pathmorphing}

\begin{document}

\title{From Kardar-Parisi-Zhang scaling to soliton proliferation in Josephson junction arrays}

\author{Mikheil Tsitsishvili}
\affiliation{Institut f\"ur Theoretische Physik, Heinrich-Heine-Universit\"at, D-40225  D\"usseldorf, Germany}
\author{Reinhold Egger}
\affiliation{Institut f\"ur Theoretische Physik, Heinrich-Heine-Universit\"at, D-40225  D\"usseldorf, Germany}
\author{Karsten Flensberg}
\affiliation{Center for Quantum Devices, Niels Bohr Institute,
University of Copenhagen, DK-2100 Copenhagen, Denmark}
\author{Sebastian Diehl}
\affiliation{Institut f\"ur Theoretische Physik, Universit\"at zu K\"oln, Z\"ulpicher Stra{\ss}e 77, D-50937 Cologne, Germany}

\begin{abstract} 
We propose Josephson junction arrays as realistic platforms for observing nonequilibrium scaling laws characterizing the Kardar-Parisi-Zhang  (KPZ) universality class, and space-time soliton proliferation.  Focusing on a two-chain ladder geometry, we perform numerical simulations for the roughness function.  Together with analytical arguments, our results predict KPZ scaling   at intermediate time scales, extending over sufficiently long time scales to be observable, followed by a crossover to the asymptotic long-time regime governed by soliton proliferation.
\end{abstract}
\maketitle

\emph{Introduction.---}The celebrated KPZ universality class, originally devised to describe the stochastic growth of interfaces \cite{Kardar1986}, covers a wide variety of nonequilibrium scaling phenomena such as the spreading of firefronts fueled by oxygen or the growth of bacterial colonies powered by sugar consumption.  KPZ scaling has been observed in various one-dimensional (1D) systems such as liquid crystals driven by electric fields \cite{Takeuchi2018}, synthetic magnets based on ultracold atoms \cite{Wei2022} or quantum information processing platforms \cite{Keenan2023}, and in polariton condensates \cite{Fontaine2022}.  We here propose Josephson junction array (JJA) setups \cite{Ustinov1998,Fazio2001,Bottcher2018,Sheikhzada2019,Fazio2021,Maffi2024} 
as highly tunable and readily available experimental platforms for probing KPZ physics, as well as space-time  defects (solitons) \cite{He2017,Moroney2023}, which imply an additional universal nonequilibrium scaling regime.  We focus on the 1D case, but as a matter of principle, JJAs also allow one to study such scaling phenomena in higher dimensions. 

\begin{figure}
    \centering
    \includegraphics[width=0.79\linewidth]{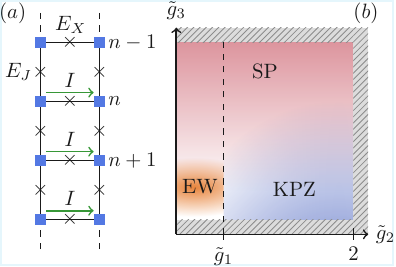}
    \caption{(a) Sketch of a JJA ladder with two 1D chains ($A$ and $B$) of $N$ superconducting grains (blue squares), $n=1,\ldots,N$.  Neighboring grains are connected via Josephson junctions (crosses), with Josephson coupling $E_J$ ($E_X$) along (across) the chains. By attaching leads to each superconducting grain, the current $I$ is driven through every rung across the $E_{X}$ junction. (b) Schematic scaling regimes in the $\tilde g_2$--$\tilde g_3$ plane for the roughness function ${\cal D}_l(t)$ in Eq.~\eqref{Ddef} at intermediate times. The dimensionless couplings $\tilde{g}_{1,2,3}$ entering our continuum model are defined below Eq.~(\ref{eq:coupling_constants}), where $\tilde g_1\propto E_X$ encodes the nonlinearity strength, $\tilde{g}_{2}\propto I$ the drive current, and $\tilde g_3\propto \sqrt{TR}$ the noise level.  In the gray dashed regions, our theory does not apply, either because superconductivity is lost (top region), the Markovian noise assumption breaks down (bottom region), or the critical current is exceeded,  $|I|>I_c=2eE_{J}/\hbar$ (right region, $\tilde g_2>2$).
    Blue regions correspond to KPZ scaling regimes spanning at least one decade in time, where more intense color implies a longer KPZ time window.  Similarly, red and orange regions indicate soliton proliferation (SP) and diffusive (EW) regimes, respectively. In the narrow white region, $k_{B}T \ll 2E_{J}$ and ${\cal D}_l(t)$ is constant.  The dashed vertical line separates regions with local minima ($\tilde g_2<\tilde g_1$) of the tilted washboard potential associated to Eq.~\eqref{langevin} from  those without minima.}
    \label{fig1}
\end{figure}

Specifically, we study the JJA ladder geometry sketched schematically in Fig.~\ref{fig1}(a), where the dynamic field $\varphi(t,x)$ is the superconducting phase difference between two chains as a function of time and space in the long wavelength continuum limit. The basic large-scale phenomenology obtained from numerical simulations and analytical theory is as follows, see also Fig.~\ref{fig1}(b). We find three successive  scaling regimes with ${\cal D}_l\propto t^{2\beta}$ for $t>0$, where ${\cal D}_l$ is the bulk roughness function defined in Eq.~\eqref{Ddef} below.  Related scaling features also appear in other observables. (i) At relatively short times, $t<t_1$, a conventional Edwards-Wilkinson (EW)  regime corresponding to free diffusive behavior with scaling exponent $\beta^{(i)}=1/4$ is observed. (ii) At intermediate time scales, $t_1<t<t_2$, typically spanning two decades, a KPZ scaling regime \cite{Kardar1986,Krug1989,Rost1994,Beijeren2012,Kulkarni2013,Tauber2014,Altman2015,Diessel2022}  characterized by anomalous diffusion with $\beta^{(ii)}=1/3$ emerges.  (iii) At asymptotically long times, $t>t_2$, space-time defects proliferate \cite{He2017,Sieberer2016,Wachtel2016,Moroney2023} and are responsible for a third regime with exponent $\beta^{(iii)}=1/2$. Snapshots of the time-dependent phase profile $\varphi(t,x)$ can directly visualize these defects. The above scenario can be experimentally probed in realistic JJA setups, offering the unique chance to observe both the KPZ and soliton proliferation regimes in the very same device under nonequilibrium conditions. Below we derive these results, estimate the time scales $t_1$ and $t_2$, and elucidate the underlying physics.

\textit{Setup, model, and observables.---}The setup in Fig.~\ref{fig1}(a) consists of two 1D JJA chains ($A$, $B$), each composed of $N$ superconducting grains with self-capacitance $C_0$. 
We assume that the grain size is large enough such that finite-size quantization effects \cite{Ralph1995,Black1996,Ralph1997,Gladilin2002} are negligible.
In contrast to the single-chain case, the ladder geometry allows one to independently tune the strength of the sine nonlinearity and the diffusion term in Eq.~\eqref{langevin} below. 
Nearest-neighbor grains are connected by the intra-chain Josephson coupling $E_J$ and by a small inter-chain (transverse) Josephson coupling $E_X\ll E_J$. We study the limit of small junction charging energies and neglect
them below. Each superconducting island in Fig.~\ref{fig1}(a) is connected to normal leads such that a constant electric current $I$ can be driven along the $E_{X}$ rung junctions. Magnetic fields generated by such currents are tiny and can be neglected in practice; for $I\approx 100$~nA, we estimate $|B|\ll 1~ \mu$T in the tunnel contact regions.
If the JJA is deposited on a two-dimensional electron gas (2DEG), current flow between the grains and the 2DEG causes Ohmic dissipation for the superconducting phase fields \cite{Fazio2001,Weiss_2012,Altland2010}.  The damping strength $\eta$ depends on the resistance $R$ between a grain and the 2DEG, which is affected by, e.g.,
the thickness of the insulating layer between the JJA and the 2DEG and/or by the chemical potential of the 2DEG. 
For now, we  assume that all junctions and/or grains have identical parameters,  but we address the impact of inhomogeneities later on.

Let us then turn to the effective model accounting for this situation. At low energies and on length scales well above the intra-chain lattice spacing $a$, using the continuous coordinate $x$, we define the local superconducting phase difference field $\varphi(t,x)=\varphi_A(t,x)-\varphi_B(t,x)$ across the chains. Since in typical experimental setups, the self-capacitance is $C_0\approx 10^{-14}$~F, the inertial term $\propto\ddot{\varphi}$ in the equation of motion for $\varphi$ can be neglected in practice. The dynamics of the JJA ladder in Fig.~\ref{fig1}(a) is then described by a driven and overdamped stochastic sine-Gordon equation (see End Matter and Refs.~\cite{Krug1989,Landauer1979}) 
\begin{equation}\label{langevin}
    \eta\dot{\varphi}-D \partial^{2}_{x} \varphi + 2E_X \sin(\varphi) - \alpha I = \xi
\end{equation}
with $\alpha = \hbar/2e$, $\eta=\alpha^{2}/R$, the diffusion constant $D=a^2E_J$, and the driving current $I$. 
The noise field $\xi(t,x)$ obeys Gaussian statistics with $\langle \xi \rangle=0$ and $\langle\xi(t,x)\xi(t',x')\rangle=4a\eta k_{B}T\delta(t-t')\delta(x-x'),$
assuming that temperature $T$ is below the superconducting critical temperature.  However, $k_B T$ should exceed the energies of the dominant 2DEG modes responsible for damping \cite{Fazio2001,Weiss_2012} such that the noise is effectively Markovian. 

As key observable probing nonequilibrium scaling in this system, we
consider the bulk roughness function,
\begin{equation}\label{Ddef}
    {\cal D}_{l}(t)=\frac{1}{l} \int_{(L-l)/2}^{(L+l)/2} dx \left( \langle \varphi^{2}(t,x)\rangle - \langle\varphi(t,x)\rangle^{2} \right),
\end{equation}
probing the time-dependent phase fluctuations averaged over a central segment of length $l<L=Na$ in order to minimize boundary effects.
Through the second Josephson relation, the local time-dependent transverse voltage is given by $V(t,x)=\frac{\hbar}{2e}\dot{\varphi}(t,x)$.  By measuring the voltage fluctuations $\langle V(t,x) V(0,x) \rangle$ and averaging over 
many noise realizations and over the spatial region of size $l$,
the roughness function \eqref{Ddef} is experimentally accessible. For details, see the End Matter.
The scaling regimes discussed above are also observable in other quantities, e.g., in the correlation function of $e^{i\varphi(t,x)}$. However, since Eq.~\eqref{Ddef} seems more accessible experimentally, we here focus on ${\cal D}_{l}$. Key to this physics are the simultaneous presence of damping, noise and diffusion terms as well as the sine nonlinearity and a driving current in Eq.~\eqref{langevin}.  Their complex interplay is realized in the JJA ladder in Fig.~\ref{fig1}(a). It is convenient to reformulate Eq.~\eqref{langevin} in terms of a dimensionless time variable $s=t/\tau$ and dimensionless couplings
\begin{equation}\label{eq:coupling_constants}
    \begin{split}
        g_{0} = E_{J}\tau R/\alpha^{2},\quad g_{1} =  2E_{X}\tau R/\alpha^{2},
        \\g_{2} = I \tau R/\alpha, \quad g_{3} = 2 \sqrt{k_{B}T \tau R} / \alpha,
    \end{split}
\end{equation}
where $\tau$ is an arbitrary reference time. Here, $g_{0},g_{1},g_{2}$ and $g_{3}$ refer to the diffusion constant, the sine nonlinearity strength, the driving term, and the noise level, respectively, while the dimensionless damping constant equals $1$.  As $\tau$ is arbitrary, we have only three dimensionless physical parameters, e.g., $\tilde g_{i=1,2,3}=g_i/g_0$.

\begin{figure}[t]
    \centering    \includegraphics[width=\linewidth]{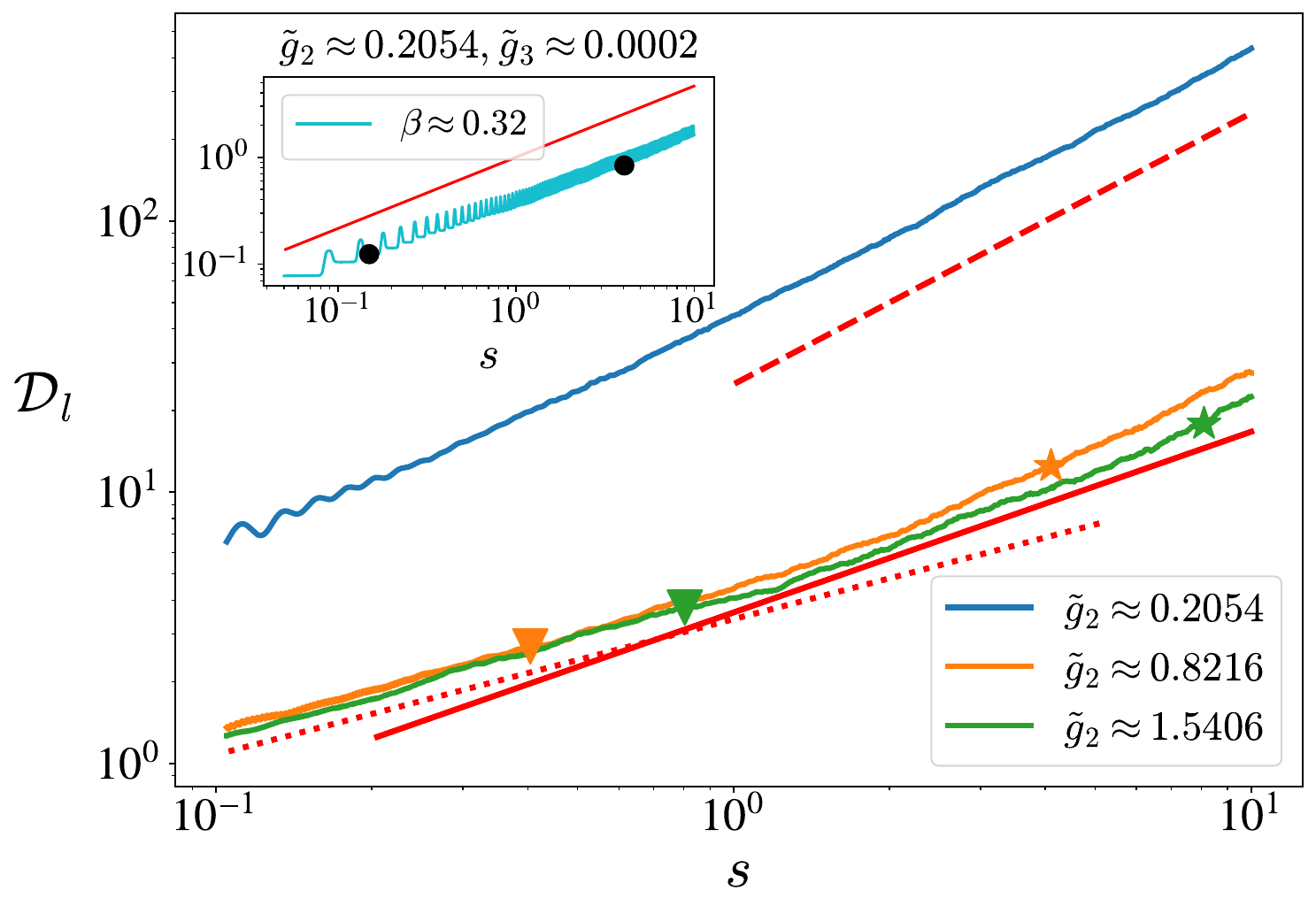}
    \caption{
    Numerical results for the roughness function ${\cal D}_{l}$ in Eq.~\eqref{Ddef} vs dimensionless time $s$ for a JJA ladder with $N=256$ grains per chain, $l= L/2$, $\tilde g_1=0.2$, and several $\tilde g_{2,3}$, see Eq.~\eqref{eq:coupling_constants}. Note the double-logarithmic scales. Red lines show the expected scaling ${\cal D}_l\sim s^{2\beta}$ for the diffusive ($\beta^{(i)}=1/4$, dotted), KPZ ($\beta^{(ii)}=1/3$, solid), and soliton proliferation regimes ($\beta^{(iii)}=1/2$, dashed), respectively. 
    Main panel: Results for $\tilde g_3\approx 0.0076$ and three values of $\tilde g_2$, with time-averaging window $\Delta s=0.01$. 
    Triangles and stars mark the approximate crossover times $s_{1}$ and $s_{2}$, respectively. Inset: Results for $\tilde{g}_{2}\approx 0.2054$ and $\tilde{g}_{3} \approx 0.0002$, with $\Delta s=0.1$, where the KPZ regime persists for more than two decades. Circles mark the time window used for extracting $\beta$.}
    \label{fig2}
\end{figure}

\emph{Numerical results.---}For numerical simulations of Eq.~\eqref{langevin} in dimensionless form (see End Matter), 
we assume the initial condition $\varphi(0,x)=0$ and impose open boundary conditions at the
chain ends.  Physically, open boundary conditions correspond to having tunnel contacts between the normal electrodes and the array structure.  Since we extract ${\cal D}_l(t)$ only in the central section of the array, however, the precise boundary condition does not affect our results for the scaling behavior. In addition, our results are robust under changes of the initial condition. In particular, if $\varphi(0,x)$ is randomly taken from a normal distribution of width $\sigma_\varphi\le 0.5$, we find identical scaling laws. We use Euler's method, reverting to discrete spatial coordinates $x\to na$ \footnote{The compactness of $\varphi_{n}$ is accounted for by letting $a^2\partial^{2}_{x}\varphi(x) \to \sum_{m=n\pm1} \sin\left(\varphi_{m}-\varphi_{n}\right)$ \cite{Sieberer2016a,He2017}.} and also discretizing time using the step size $\delta s=10^{-4}$.  We then compute ${\cal D}_l(s)$ from Eq.~\eqref{Ddef} and average over $1000$ noise trajectories.
The size of the central averaging region in Eq.~\eqref{Ddef} was chosen as $l=L/2$, but 
we have checked that using smaller $l$ (down to $l=L/16$) gave the same 
power-law exponents $\beta$ in the respective regimes.   
Numerical results for ${\cal D}_l(s)$ for various temperatures and/or currents are shown in Fig.~\ref{fig2}, where we find all three scaling regimes discussed above.
For all parameters in Fig.~\ref{fig2}, we have $\tilde{g}_{2} > \tilde{g}_{1}$, where the
tilted washboard potential corresponding to Eq.~\eqref{langevin} has no local minima.  To extract power-law scaling exponents, the resulting oscillations in ${\cal D}_l(s)$ were suppressed by temporal averaging over a time window of size $\Delta s$.
In all cases below, these exponents were obtained by fitting numerical data for ${\cal D}_l(s)$ to the function $\tilde{a} s^{2\tilde{\beta}}$ (with fitting parameters $\tilde{a}$ and $\tilde{\beta}$) over an extended time range.

The main panel of Fig.~\ref{fig2} shows results for $E_{J}=200~\mu$eV, $E_{X}=20\mu$eV, $R=0.1~\Omega$, $T=0.1$~K, using $I=0.02$, $0.08$ and $0.15~\mu$A, where $\delta t=10$~ns corresponds to $\delta s=10^{-4}$.
Using Eq.~\eqref{eq:coupling_constants}, we specify the corresponding dimensionless couplings $\tilde g_{1,2,3}$ in Fig.~\ref{fig2}. For weak driving current,  $\tilde g_2\approx 0.2054$, only the soliton proliferation regime with $\beta^{(iii)}\approx 0.5$ is observed. However, the intermediate KPZ regime emerges for larger driving currents. Indeed, for $\tilde{g}_{2} \approx 0.8216$, we find good agreement with the diffusive power-law exponent $\beta^{(i)}\approx 0.26$ for short times $s<s_{1}$, while the KPZ  exponent  $\beta^{(ii)}\approx 0.37$ emerges at intermediate times $s_{1}<s<s_2$. At long times, one approaches the soliton proliferation regime, $\beta^{(iii)}\approx 0.47$ for $s>s_2$.  Finally, for the largest current with $\tilde g_2\approx 1.5406$, the KPZ regime extends over more than a decade in time. Our numerical fit now gives $\beta^{(i)}\approx 0.26$ for $s<s_{1}$, $\beta^{(ii)}\approx 0.37$ for $s_{1}<s<s_2$, and $\beta^{(iii)}\approx 0.51$ for $s>s_2$.  
We note in passing that the onset time $s_2$ for the soliton proliferation regime becomes shorter with increasing temperature. 
Next, the inset of Fig.~\ref{fig2} shows results for $E_{J} = 2$~meV, $E_{X} = 0.2$~meV, $R=0.01~\Omega$, $I = 0.2~\mu$A, and $T=1$~mK. Now a wide KPZ scaling regime extending over more than two decades is observed, with $t_1\approx 10$~$\mu$s and $t_2\approx 1$~ms in physical units. 

\begin{figure}
    \centering
    \includegraphics[width=\linewidth]{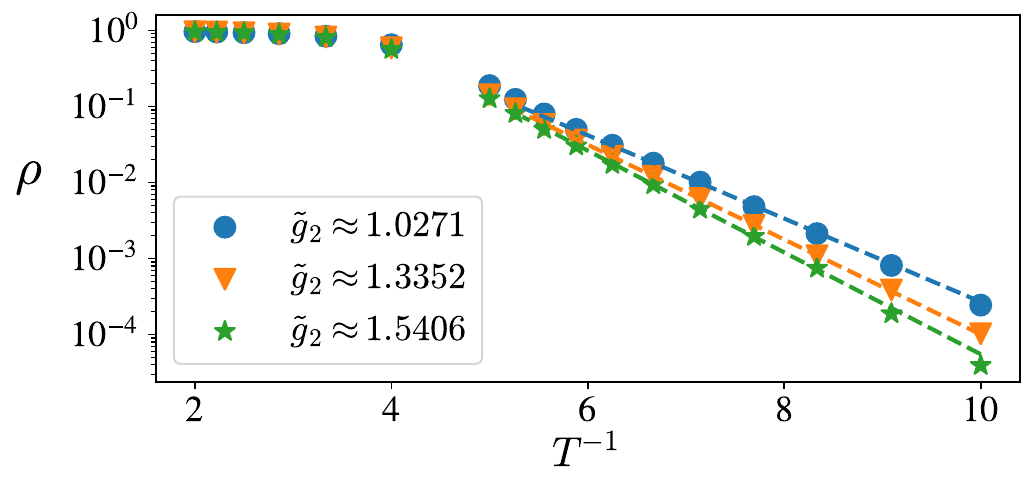}
    \caption{Numerical results for the soliton density $\rho$ (in units of $1/a$) vs inverse temperature ($T$ in Kelvin) in the overdriven regime $\tilde g_2>\tilde g_1=0.2$ for several $\tilde g_2$ on a log-linear scale.  For small noise levels $\tilde g_3$ corresponding to low $T$, see Eq.~\eqref{eq:coupling_constants}, we find activated behavior, $\rho \sim \exp(-E_A/k_B T)$. Numerical fits (dashed lines) yield $E_{A}\approx 109, 124$ and $133~\mu$eV with increasing $\tilde{g}_{2}$, respectively.}
    \label{fig3}
\end{figure}

Let us now consider the field $\Theta(s,x) \equiv \varphi(s,x) - \varphi_{0}(s,x)$
for a representative single noise trajectory $\varphi(s,x)$, where $\varphi_{0}(s,x)$ is the corresponding noiseless solution of Eq.~(\ref{langevin}) in dimensionless units. By numerically calculating $\partial_{x}\Theta(s,x)$ for every time step, we observe that $\partial_x\Theta(s,x) \approx 2\pi m$ with $m\in \mathbb{Z}$ is always fulfilled. Locations with $m\ne 0$ are identified as the space-time solitons. 
Since we use open boundary conditions, regions near the chain edges favor such defect formation. To obtain results independent of these boundary effects, we use $l=L/2$ in Eq.~\eqref{Ddef}. The problem of thermally activated solitons in sine-Gordon models has previously been studied in the ``underdriven'' limit $\tilde g_2<\tilde g_1$, where the washboard potential has well-defined stable minima~\cite{Landauer1979,Fitzgerald2016,Kolton2020}. For the ``overdriven'' case 
$\tilde{g}_2 > \tilde{g}_{1}$, however, there are no minima and $\varphi(s,x)$ slides down indefinitely, even for $\tilde{g}_{3} = 0$. Nonetheless, by coarse-graining time, below we construct an effective theory with well-defined space-time solitons in the overdriven regime where, in fact, our numerical results directly reveal their presence. In Fig.~\ref{fig3}, we show the soliton density $\rho$ found numerically at long times. We define $\rho=\langle N_{l}\rangle/l$, where $\langle N_{l}\rangle$ is the noise averaged number of solitons in the central segment $l$. At low temperatures, Fig.~\ref{fig3} reveals activated behavior, $\rho \sim e^{-E_A/k_B T}$ with activation energy $E_A$. At high temperatures, $k_B T\gg E_A$, solitons proliferate and their density $\rho\approx 1/a$ saturates.  

\emph{Theoretical interpretation.---}The scaling regimes (i) and (ii)  can be understood based on the spontaneous breaking of time translation symmetry, see also Ref.~\cite{Fujii2022}, while regime (iii) roots in the compactness of $\Theta(t,x) = \varphi(t,x)-\varphi_{0}(t,x)$, inherited from the periodicity of the sine nonlinearity in Eq.~\eqref{langevin}.  
We first show how broken time translations explain the scaling behavior in (i) and (ii).  
Here the key ingredient is the combination of current driving, the bounded potential $\sim E_X$, and the non-vanishing noise level. Indeed, Eq.~\eqref{langevin} describes the stochastic motion of an effective particle in a tilted washboard potential. In the overdriven case, a steady-state particle ``current'' will be running, $\langle \dot\varphi (t,x)\rangle = v \neq 0$, and hence $\langle \varphi (t,x)\rangle = v t$. For small $D$ and $E_X$, we estimate $v\approx 2e RI/\hbar$, see also Ref.~\cite{Ivanchenko1969}.  By the Josephson relation, $\langle\dot\varphi\rangle=v$ could be directly measured in terms of a voltage drop along the transverse direction.  Despite Eq.~\eqref{langevin} being invariant under time translations, its solution is not -- an instance of spontaneous breaking of a continuous symmetry, which implies the existence of a gapless Goldstone mode. Clearly, this argument has to be taken with the usual grain of salt in 1D due to strong fluctuations. Nonetheless, the construction of these fluctuations allows us to make progress:
if $\varphi_{0}(t,x)$ is a solution of Eq.~\eqref{langevin} in the absence of noise, also $\varphi_0(t+t_0,x)$ is a solution at sufficiently long times such that the memory about initial conditions is lost. According to the Goldstone construction, we replace the global symmetry transformation $t_0$ by a slowly varying dynamic field, $t_0 \to \theta (t,x)$. We thus parametrize the field as $\varphi(t,x) = \varphi_0(t+\theta(t,x),x)$, where  variations of $\theta$ take place on much larger space-time scales than those of $\varphi_{0}$. 
The equation of motion for the slow mode $\theta$ is the KPZ equation \cite{Kardar1986},
\begin{align}\label{eq:kpz}
    Z \partial_t\theta - K \partial_x^2\theta +  \lambda(\partial_x\theta)^2 = \zeta,
\end{align}
where $Z=\eta \dot\varphi_{0}\approx \eta v,$  $K=D \dot\varphi_{0}\approx D v,$ $\lambda =D\ddot\varphi_{0},$ while $\varphi_{0}$ itself encodes $E_X$. The noise level of $\zeta(t,x)$ is approximately equal to that of $\xi$ in Eq.~\eqref{langevin}.
On this level, the nonlinearity $\lambda$ is small -- this is the origin of regime (i), showing diffusive scaling. Coarse graining to larger scales, however, the nonlinearity grows since it is a relevant coupling in 1D \cite{Takeuchi2018,Krug1997}. Once the scales have grown such that the dimensionless parameter $\tilde \lambda = \lambda (k_{B}T/D^3)^{1/2}$ has reached unity, the system crosses over to regime (ii) which is governed by the strong-coupling KPZ fixed point with scaling exponent $\beta^{(ii)}=1/3$ \cite{Krug1997,Calabrese2014,He2017}. For large  driving current, $|I|\gg 2E_X/\alpha$ (but $|I| < I_c$), the time scale $t_{1}$ for the onset of the KPZ regime is estimated by following Ref.~\cite{Nattermann1992} (see End Matter),
\begin{equation}   \label{t1estimate}
t_{1} \approx t_\Delta \left(\frac{E_{J}}{\epsilon^2 k_{B} T}\right)^{2}, \quad \epsilon \equiv 2E_{X}/\alpha I,
\end{equation} 
where $t_\Delta$ is a microscopic time scale.  Since the superconducting
gap $\Delta$ implies a lower limit for the time resolution in Eq.~\eqref{langevin}, we expect $t_\Delta\approx \hbar/\Delta$. 

The mechanism leading to Eq.~\eqref{eq:kpz} does not rely on continuous \textit{internal} symmetries, which are indeed absent in Eq.~\eqref{langevin}. Instead, it relies on the spontaneous breaking of the \textit{external} symmetry of time translations which is forbidden in equilibrium~\cite{Bruno2013,Watanabe2015}. It can only occur out of equilibrium and/or for unbounded generators of dynamics as realized for the tilted washboard potential. This mechanism was discussed recently for time crystals \cite{Daviet2024} (see also \cite{Cuerno2024_1,Cuerno2024_2}), where the Goldstone mode is compact, $\theta \in SO(2)$. However, $\theta$ is non-compact in our case, $\theta \in  \mathbb{R}$. 
While the mapping to the KPZ equation with non-compact variable $\theta$ explains regimes (i) and (ii), it does not account for regime (iii) which instead is caused by the compactness of $\Theta(t,x)$ introduced above.
To see this, we coarse-grain Eq.~\eqref{eq:kpz} in time and construct $\varphi_{0}(t,x)$ perturbatively in $\epsilon$, see Eq.~\eqref{t1estimate}. Up to first order, with $\omega=\alpha I/\eta$, we find 
$\varphi^{(1)}_{0} = \omega t - 2\epsilon\sin^2(\omega t/2)$.
Coarse-graining Eq.~\eqref{eq:kpz} over the time scale $t_*=2\pi/\omega$, with 
$\overline{\Theta}(t,x) = t^{-1}_{*}\int^{t+t_{*}}_{t}dt' \Theta(t',x)$,  we get
(see End Matter)
\begin{equation}\label{eq:effective_sG}
\eta \dot{\overline{\Theta}} - D\partial^{2}_{x}\overline{\Theta} - \epsilon F_{{\rm eff}}(\overline{\Theta},\dot{\overline{\Theta}}) = \overline{\xi}
\end{equation}
with an effective force $F_{{\rm eff}} = 
2E_{X}\sin^{2}(\overline{\Theta}/2) - \eta \dot{\overline{\Theta}} \sin \overline{\Theta}$.
The noise level $\overline{\xi}$ is approximately as for $\xi$ in Eq.~\eqref{langevin}. The effective force consists of the washboard potential at the depinning point and a non-conservative ``restoring damping'' force. The potential force  is always positive and vanishes only if $\overline{\Theta} = 2\pi n$ ($n\in \mathbb{Z}$). The rolling motion down the potential is initialized by thermal fluctuations with $\dot{\overline{\Theta}} > 0$, which activates a restoring damping force in the opposite direction, driving $\overline{\Theta}$ back to the original value. However, if the thermal kick is so large that $\overline{\Theta}>\pi$,
the damping force changes sign and drives one toward
$\overline{\Theta} = 2\pi (n+1)$.  The interplay of thermal fluctuations, rolling motion, and non-conservative forces thus pins $\overline{\Theta}=2\pi n$, thereby allowing for stable space-time solitons.  These solitons destroy the coherence of $e^{i\varphi(t,x)}$ at asymptotic space-time distances and lead to an exponential decay of the corresponding correlator~\cite{He2017}. This phase scrambling effect implies $\beta^{(iii)}=1/2$  for the roughness function ${\cal D}_l(t)$. 
The crossover time $t_2$ between the KPZ and soliton proliferation 
regimes can be estimated by noting that the probability to encounter a soliton is  $p\approx a\rho \approx e^{-E_{A}/k_{B}T}$, where the activation energy $E_A$ depends on the parameters in Eq.~\eqref{langevin}, see Fig.~\ref{fig3}.
We thus arrive at
\begin{eqnarray}\label{t2estimate}
   t_{2} \sim t_{\Delta} e^{E_A/k_{B}T}.
\end{eqnarray} 
The estimates for $t_1$ and $t_2$ in Eqs.~\eqref{t1estimate} and \eqref{t2estimate}, respectively, are qualitatively consistent with our numerical results. 
We note that at sufficiently high $T$, one finds $t_2\alt t_1$ and the KPZ regime (ii) disappears. On the other hand, for low $T$, one can achieve a wide window $t_2-t_1$ where  KPZ scaling is observable.  We emphasize that both the model and the physical mechanisms behind the various scaling regimes differ qualitatively and quantitatively from previous work for exciton-polariton systems, see, e.g., Ref.~\cite{He2017}.  In particular, the solitons described above for the overdriven regime are distinct from the conventional sine-Gordon solitons appearing in the underdriven regime. 

\emph{Discussion.---}We have also studied effects of spatial inhomogeneities in the parameters $(E_J,E_X,R)$ along the ladder direction. From simulations for samples drawn from a uniform random distribution with width $\pm 2\%$ for each parameter centered around the values in Fig.~\ref{fig2}, we observe that disorder favors soliton formation and narrows down the KPZ scaling window. However, a reduction of temperature by a factor of two is sufficient to suppress soliton generation and recover a full decade of the KPZ regime. Since $(E_J,E_X,R)$ can be adjusted to accuracy better than $2\%$ in available JJAs \cite{Rasmussen2021}, sample inhomogeneities should not be detrimental to observing the predicted dynamical scaling laws. Furthermore, in a transverse magnetic field, defining $f=\Phi/\Phi_0$ with the magnetic flux $\Phi$ through  a plaquette and $\Phi_0=h/2e$, a pitch $\sin\left(\varphi+2\pi f x/a\right)$ appears in the sine nonlinearity. This leads to surprisingly rich and  complex features to be discussed elsewhere.  Moreover, generalizations to 2D geometries are highly desirable since there is scarce experimental evidence for KPZ scaling \cite{Ojeda2000,Almeida2014}, with driven quantum matter platforms -- polaritons -- entering the game only very recently~\cite{widmann2025}, and nonequilibrium effects play an even more fundamental role than in 1D.

\emph{Data availability.---}All data underlying the figures in this paper are available at Zenodo \cite{Zenodo}.

\begin{acknowledgments} 
We thank B. Beradze, M. Burrello, R. Daviet, R. Fazio, J. Krug, C.~M.~Marcus, A. Rosch, S.~Vaitiek{\.e}nas, C.~P. Zelle, and especially T. Giamarchi for discussions. We acknowledge funding by the Deutsche Forschungsgemeinschaft (DFG, German Research Foundation) under Projektnummer 277101999 - TRR 183 (projects B02 and C01), under Projektnummer EG 96/14-1, and under Germany's Excellence Strategy - Cluster of Excellence Matter and Light for Quantum Computing (ML4Q) EXC 2004/1 - 390534769.
\end{acknowledgments}

\bibliography{biblio}

\appendix
\section*{End Matter}
\setcounter{equation}{0}
\setcounter{figure}{0}
\renewcommand{\thefigure}{A\arabic{figure}}
\renewcommand{\theequation}{A\arabic{equation}}

We here provide derivations for expressions given in the main text and additional details. 

First, we outline the derivation of Eq.~\eqref{langevin}.   Using the Schwinger-Keldysh approach
\cite{Altland2010}, we define superconducting phase fields $\phi_{n,\alpha,\sigma}(t)$ for the $n$th grain pertaining to chain 
$\alpha\in\{A,B\}$, where $\sigma=\pm$ refers to the forward and backward part of the Keldysh time contour.  Exploiting that one can expand the cosine in $E_J\cos(\phi_{n,\alpha,\sigma}-\phi_{n+1,\alpha,\sigma})$ for large $E_J$,
the action is given by ($t_f\to \infty$ is the final time on the Keldysh contour)
\begin{eqnarray}\nonumber
S &=& S_{\rm diss} + S_I + \int_0^{t_f} dt \Biggl(\frac{\hbar^{2} C_0}{8e^{2}} \sum_{n,\alpha,\sigma}\sigma \dot\phi^2_{n,\alpha,\sigma} \\ \label{keldact}
&+& \sum_{\langle n,n'\rangle,\alpha,\sigma} \frac{\sigma E_J}{2} \left[\phi_{n,\alpha,\sigma}-
\phi_{n',\alpha, \sigma}\right]^2 \\
\nonumber &-& E_X\sum_{n,\sigma} \sigma\cos[\phi_{n,A,\sigma}-\phi_{n,B,\sigma}]   \Biggr), 
\end{eqnarray}
where the applied current $I$ enters through  
\begin{equation}
    S_I =  \sum_{n,\sigma} \frac{\sigma \hbar I}{2e} \int_0^{t_f} dt \left(\,\phi_{n,A,\sigma}-\,\phi_{n,B,\sigma} \right).
\end{equation}
The coupling of each grain to the underlying 2DEG gives rise to a dissipative term $S_{\rm diss}$, 
which can be written in standard Feynman-Vernon form as familiar from Caldeira-Leggett models
\cite{Fazio2001,Weiss_2012}.  Assuming that 2DEG modes coupling to different grains are uncorrelated,
and introducing ``classical'' (cl) and ``quantum'' (qu) fields as $\phi_{n,\alpha}^{\rm (cl)}
= \frac12 \sum_\sigma \phi_{n,\alpha,\sigma}$ and $\phi_{n,\alpha}^{\rm (qu)}=\sum_\sigma\sigma\phi_{n,\alpha,\sigma}$,
we arrive at 
\begin{eqnarray}\nonumber
    S_{\rm diss} &=& i\sum_{n,\alpha}\int_0^{t_f} dt\int_0^t dt'\, L(t-t')
    \phi_{n,\alpha}^{\rm (qu)}(t)
    \phi_{n,\alpha}^{\rm (qu)}(t') \\ \label{dissact}  
    &-& \eta \sum_{n,\alpha} \int_0^{t_f} dt\, \phi_{n,\alpha}^{\rm (qu)}(t) \dot\phi_{n,\alpha}^{\rm (cl)}(t),
\end{eqnarray}
with the bath correlation function $L(t)=\frac{1}{\pi}\int_0^\infty d\omega\, J(\omega)
\coth\left(\frac{\hbar\omega}{2k_{B}T}\right) \cos(\omega t)$.  We assume an Ohmic spectral density, $J(\omega)=\eta\omega e^{-\omega/\omega_c}$, up to frequencies of order $\omega_c$.  
Since we consider relatively high temperatures, $k_{B}T\agt \hbar\omega_c$,
we can approximate $L(t)\simeq 2\eta k_{B}T\delta(t)/\hbar$ by a Markovian correlator.
Similarly, we may take the semiclassical limit and expand in the ``quantum'' fields $\phi_{n,\alpha}^{\rm (qu)}$ capturing quantum fluctuations.  The dissipative action can then equivalently be 
expressed in terms of Gaussian noise fields $\xi_{n,\alpha}(t)$ with zero mean and the correlator
$\langle \xi_{n,\alpha}(t) \xi_{n',\alpha'}(t')\rangle = 2 \eta k_{B}T \delta_{n,n'} \delta_{\alpha,\alpha'} \delta(t-t').$
As a consequence, one arrives at Langevin equations for the classical fields $\phi_{n,\alpha}^{\rm (cl)}(t)$ \cite{Weiss_2012}.  Defining  phase differences, $\varphi_n(t)=\phi_{n,A}^{\rm (cl)}(t)-\phi_{n,B}^{\rm (cl)}(t)$, and the sum of  phases, $\vartheta_n(t)=\frac12\sum_\alpha \phi_{n,\alpha}^{\rm (cl)}(t)$, the equations for $\varphi_n$ and $\vartheta_n$ decouple.  Since the sine nonlinearity $\sim E_X$ only shows up in the equation for $\varphi_n$, 
we focus on the latter equation.  
Taking the continuum limit, $na\to x$, and neglecting the inertial term due to $C_0$, we then arrive at Eq.~\eqref{langevin}, where
the noise field $\xi(t,x)$ follows
from $\xi_{n,A}(t)-\xi_{n,B}(t)$,  with the correlator specified in the main text 
after a rescaling $\eta\to 2\eta$ to simplify notation.  Similarly, the diffusion term in Eq.~\eqref{langevin} 
is obtained from $\varphi_{n+1}+\varphi_{n-1}-2\varphi_n\to a^2\partial_x^2
\varphi(x)$. We note that one can alternatively derive 
Eq.~\eqref{langevin} from the circuit equations for the resistively shunted setup in Fig.~\ref{fig1}(a) \cite{Fazio2001,Fazio2021}. After expanding the along-chain Josephson term to lowest nontrivial order and including the Johnson-Nyquist noise due the resistors, one eventually also arrives at Eq.~\eqref{langevin}. 

Let us now elaborate on the measurement protocol for the roughness function ${\cal D}_l(t)$ in Eq.~\eqref{Ddef}. We start by defining the fluctuation correlator of the phase variable, spatially averaged over the central region of size $l$ and averaged over many noise realizations,
\begin{equation}\label{Ddef2}
    \begin{split}
        \Delta_{\varphi}(t_{1}-t_{2}) = \frac{1}{l}\int dx &\big(\langle [\varphi(x,t_{1}) - \varphi(x,t_{2})]^{2} \rangle 
        \\
        - &\langle \varphi(x,t_{1}) - \varphi(x,t_{2}) \rangle^{2} \big).
    \end{split}
\end{equation}
With the local voltage $V(t,x)$ introduced in the main text, we get
\begin{equation}
    \begin{split}
        \frac{\partial^{2}\Delta_{\varphi}}{\partial t_{2}\partial t_{1}} \propto \frac{1}{l}\int dx &\big(\langle V(x,t_{1})\rangle \langle V(x,t_{2})\rangle 
        \\
        -&\langle V(x,t_{1}) V(x,t_{2})\rangle \big) \equiv \mathcal{V}(t_{1}-t_{2}),
    \end{split}
\end{equation}
defining the averaged time-dependent voltage correlations ${\cal V}(t_1-t_2)$. For instance, in the KPZ time window, one expects the scaling $\Delta_{\varphi} \propto |t_{1}-t_{2}|^{2/3}$, and thus $\mathcal{V}(t_{1}-t_{2}) \propto |t_{1}-t_{2}|^{-4/3}$. Without loss of generality, we may put $t_2=0$ in Eq.~\eqref{Ddef2} and use the initial condition $\varphi(x,0)=0$.  The resulting expression for $\Delta_\varphi(t)$ coincides with ${\cal D}_l(t)$ in Eq.~\eqref{Ddef}.  We conclude that by measuring the voltage noise correlations, one can experimentally access the scaling regimes identified in the main text. 

Next, by using Eq.~\eqref{eq:coupling_constants} and the dimensionless time variable $s=t/\tau$, we express Eq.~\eqref{langevin} in dimensionless units,
\begin{equation}\label{langevin-dimensionless}
    \partial^{\phantom{2}}_{s} \varphi -g_{0}\partial^{2}_{x} \varphi + g_{1}\sin(\varphi) - g_{2} = g_{3}\tilde{\xi}.
\end{equation}
The rescaled noise field $\tilde{\xi}$ satisfies $\langle\tilde{\xi}(s,x)\tilde{\xi}(s',x')\rangle = a\delta(s-s')\delta(x-x')$. Moreover, Eqs.~(\ref{langevin-dimensionless}) and (\ref{eq:coupling_constants}) stay invariant under the scaling transformation
\begin{eqnarray}\label{lambda-transformation}
    \{E_{J},E_{X},I,T,R\} \to \{\nu E_{J},\nu E_{X},\nu I,\nu T,R/\nu \}
\end{eqnarray}
for arbitrary real number $\nu>0$. A solution of Eq.~\eqref{langevin-dimensionless} for  given coupling set $\{g_{0,1,2,3}\}$ thus immediately provides a one-parameter manifold of solutions to Eq.~\eqref{langevin}.

\begin{figure}
    \centering
    \includegraphics[width=1.0\linewidth]{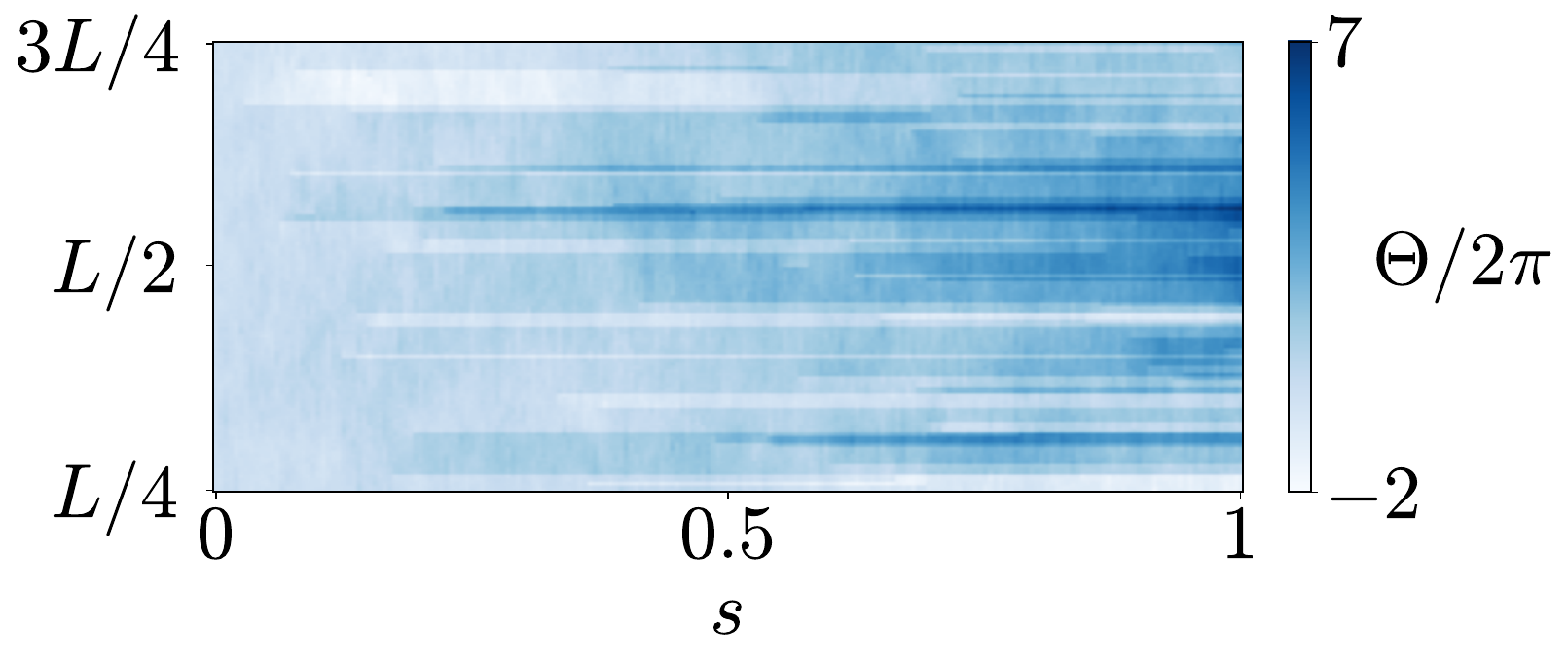}
    \caption{Color-scale plot of $\Theta(s,x) = \varphi(s,x)-\varphi_{0}(s,x)$ in the space-time plane restricted to a central segment $l=L/2$ of the ladder. We consider the soliton proliferation regime with $\tilde g_1=0.2$, $\tilde g_2\approx 1.027$ and $\tilde g_3\approx 0.0132$, where $\varphi(s,x)$ and $\varphi_{0}(s,x)$ are numerical solutions of Eq.~(\ref{langevin-dimensionless}), with $\varphi$ obtained for a single noise trajectory and $\varphi_0$ without noise ($\tilde g_3=0$). Space-time solitons are identified as explained in the main text.}
    \label{figA1}
\end{figure}

Second, we turn to a derivation of Eq.~\eqref{eq:kpz} and the corresponding onset time $t_{1}$ for the KPZ regime.  Writing $\varphi(t,x) = \varphi_0(t+\theta(t,x),x)$, where $\varphi_0$ is the solution of Eq.~\eqref{langevin} in the absence of noise and using $t' \equiv t+\theta(t,x)$, we arrive at 
\begin{eqnarray}
        \partial_{t} \varphi(t,x)&=& \partial_{t'}\varphi_{0}(t',x) \, \left[1+\partial_{t} \theta(t,x)\right],
        \\ \nonumber
        \partial^{2}_{x}\varphi(t,x)&=&\partial^{2}_{x} \varphi_{0}(t',x)+\partial_{t'} \varphi_{0}(t',x) \, \left[ \partial^{2}_{x} \theta(t, x) \right]
        \\ \nonumber
        &+&\partial^{2}_{t'} \varphi_{0}(t',x) \, \left[\partial_{x} \theta(t, x)\right]^2.
\end{eqnarray}
Plugging these expressions into Eq.~\eqref{langevin}, we find that the dynamics of the slow 
field $\theta(t,x)$ is governed by the KPZ equation \eqref{eq:kpz}. 
By rescaling the field $\theta = \sqrt{4\eta^2k_{B}T/DZ^2} \,\tilde\vartheta$ and time $t= (\eta/D)\tilde t$, we get from Eq.~\eqref{eq:kpz} the equation of motion for $\tilde \vartheta(\tilde t,x)$:
\begin{equation}
        \partial_{\tilde t} \tilde\vartheta - \partial^2_x \tilde\vartheta+ 
         g(\partial_x \tilde\vartheta)^2=  \tilde{\xi}(\tilde t,x),
\end{equation}
where $g = 2\sqrt{k_{B}T/D}\, \ddot{\varphi}_{0}/\dot{\varphi}_{0}^{2}$ and the noise level $\tilde\xi$ has the  correlator $\langle \tilde{\xi}(\tilde t,x) \tilde{\xi}(\tilde t',x')\rangle = a \delta(\tilde t -\tilde t')\delta(x-x')$. Going back to Eq.~\eqref{langevin} for $\varphi_{0}$, we assume a flat initial condition where
no spatial gradient emerges. With $\epsilon$ in Eq.~\eqref{t1estimate}, we then obtain 
 $\ddot{\varphi}_{0}/\dot{\varphi}_{0}^{2} = \epsilon\cos(\varphi_{0})/[\epsilon\sin(\varphi_{0}) - 1]$, see also Ref.~\cite{Ivanchenko1969}.
For $\epsilon \ll 1$, we find $\left| \ddot{\varphi}_{0}/\dot{\varphi}_{0}^{2} \right| \lesssim \epsilon$.
Since $t_{1} \sim |g|^{-4}$ \cite{Nattermann1992}, we arrive at Eq.~\eqref{t1estimate}. In particular,  one can extend $t_{1}$ to longer times by reducing $|g|$.  

Finally, we discuss the overdriven regime $\tilde g_2>\tilde g_1$.
Numerical evidence for the emergence of space-time solitons in the overdriven regime is shown in Fig.~\ref{figA1}, where $\Theta(s,x)$ exhibits transitions between the pinned values $2\pi n$ with integer $n$.
The equation of motion for the field 
$\Theta(t,x)=\varphi(t,x)-\varphi_{0}(t,x)$ is given by 
\begin{equation}\label{eq:full-eq-Theta}
\eta\dot{\Theta} - D\partial^{2}_{x}\Theta +2E_{X}\left[\sin(\varphi_{0} + \Theta) - \sin(\varphi_0)\right] = \xi.
\end{equation}\\
Coarse-graining Eq.~\eqref{eq:full-eq-Theta} over $t_{*}=2\pi/\omega$, we get
\begin{equation}
    \eta\dot{\overline{\Theta}} - D\partial^2_{x}\overline{\Theta}+2E_{X}\left(\overline{\sin(\varphi_{0} + \Theta)} - \overline{\sin(\varphi_0)}\right)= \overline{\xi}
\end{equation}\\
with $\overline{\sin(\varphi_{0} + \Theta)}$ and $\overline{\sin(\varphi_{0})}$ denoting time averages over the time window $t_{*}$. These averages follow as
\begin{equation}\label{eq:coarse-graining}
 \overline{f\left(\varphi_{0}(t)\right)g\left(\Theta(t)\right)} = \sum^{\infty}_{n=0}\frac{F_{n}(t)}{n!}\frac{d^{n}}{dt^{n}}g\left(\Theta(t)\right)
\end{equation}
with $F_{n}(t) = \frac{1}{t_{*}}\int^{t_{*}}_{0}f\left(\varphi_{0}(t+\tau)\right) \tau^{n} d\tau.$  We note that Eq.~\eqref{eq:full-eq-Theta} is equivalent to Eq.~\eqref{eq:kpz}, as can be seen by inserting $\Theta(t,x)=\varphi_{0}(t+\theta(t,x),x)-\varphi_{0}(t,x)$ into Eq.~\eqref{eq:full-eq-Theta}. By using the first-order approximation $\varphi^{(1)}_{0}(t,x)$, truncating Eq.~\eqref{eq:coarse-graining} at $n=1$, and dropping rapidly oscillating terms, we arrive at the expression for $F_{\rm eff}$ quoted in the main text. 

\end{document}